\documentclass[amsmath, preprintnumbers, sort&compress, twocolumn]{revtex4}

\usepackage{graphicx,epsfig}
\usepackage{mathptmx, textcomp}
\usepackage[latin1]{inputenc}

\newcommand{\rs}{\rm \scriptscriptstyle}
\newcommand{\bs}{\bf \scriptscriptstyle}

\begin{document}
\title{Repulsively bound atom pairs in an optical lattice}





\author{K. Winkler}
\affiliation{Institute for Experimental Physics, University of Innsbruck,
A-6020 Innsbruck, Austria} \affiliation{$^1$ Institute for Quantum Optics and
Quantum Information of the Austrian Academy of Sciences, A-6020 Innsbruck,
Austria}
\author{G. Thalhammer}
\affiliation{Institute for Experimental Physics, University of Innsbruck,
A-6020 Innsbruck, Austria} \affiliation{$^1$ Institute for Quantum Optics and
Quantum Information of the Austrian Academy of Sciences, A-6020 Innsbruck,
Austria}
\author{F. Lang}
\affiliation{Institute for Experimental Physics, University of Innsbruck,
A-6020 Innsbruck, Austria} \affiliation{$^1$ Institute for Quantum Optics and
Quantum Information of the Austrian Academy of Sciences, A-6020 Innsbruck,
Austria}
\author{R. Grimm$^1$}
\affiliation{Institute for Experimental Physics, University of Innsbruck,
A-6020 Innsbruck, Austria} \affiliation{$^1$ Institute for Quantum Optics and
Quantum Information of the Austrian Academy of Sciences, A-6020 Innsbruck,
Austria}
\author{J. Hecker Denschlag}
\affiliation{Institute for Experimental Physics, University of Innsbruck,
A-6020 Innsbruck, Austria}
 \affiliation{$^1$ Institute for Quantum Optics and Quantum Information of the Austrian
Academy of Sciences, A-6020 Innsbruck, Austria}

\author{A. J. Daley}
\affiliation{Institute for Theoretical Physics, University of Innsbruck, A-6020
Innsbruck, Austria}\affiliation{Institute for Quantum Optics and Quantum
Information of the Austrian Academy of Sciences, A-6020 Innsbruck, Austria}
\author{A. Kantian}
\affiliation{Institute for Theoretical Physics, University of Innsbruck, A-6020
Innsbruck, Austria}\affiliation{Institute for Quantum Optics and Quantum
Information of the Austrian Academy of Sciences, A-6020 Innsbruck, Austria}
\author{H. P. B\"uchler}
\affiliation{Institute for Theoretical Physics, University of Innsbruck, A-6020
Innsbruck, Austria}\affiliation{Institute for Quantum Optics and Quantum
Information of the Austrian Academy of Sciences, A-6020 Innsbruck, Austria}
\author{P. Zoller}
\affiliation{Institute for Theoretical Physics, University of Innsbruck, A-6020
Innsbruck, Austria}\affiliation{Institute for Quantum Optics and Quantum
Information of the Austrian Academy of Sciences, A-6020 Innsbruck, Austria}

\date{8 May 2006}

\begin{abstract}
Throughout physics, stable composite objects are usually formed via attractive
forces, which allow the constituents to lower their energy by binding together.
Repulsive forces separate particles in free space. However, in a structured
environment such as a periodic potential and in the absence of dissipation,
stable composite objects can exist even for repulsive interactions. Here we
report on the first observation of such an exotic bound state, comprised of a
pair of ultracold atoms in an optical lattice. Consistent with our theoretical
analysis, these repulsively bound pairs exhibit long lifetimes, even under
collisions with one another. Signatures of the pairs are also recognised in the
characteristic momentum distribution and through spectroscopic measurements.
There is no analogue in traditional condensed matter systems of such
repulsively bound pairs, due to the presence of strong decay channels. These
results exemplify on a new level the strong correspondence between the optical
lattice physics of ultracold bosonic atoms and the Bose-Hubbard
model\cite{hubbardtoolbox,Blo05}, a correspondence which is vital for future
applications of these systems to the study of strongly correlated condensed
matter systems and to quantum information.
\end{abstract}

\maketitle

Cold atoms loaded into a 3D optical lattice provide a realisation of a quantum
lattice gas\cite{hubbardtoolbox,Blo05}. An optical lattice can be generated by
pairs of counterpropagating laser beams, where the resulting standing wave
intensity pattern forms a periodic array of microtraps for the cold atoms, with
period $a$ given by half the wavelength of the light, $\lambda /2$. The
periodicity of the potential gives rise to a bandstructure for the atom
dynamics with Bloch bands separated by band gaps, which can be controlled via
the laser parameters and beam configuration. The dynamics of ultracold atoms
loaded into the lowest band of a sufficiently deep optical lattice is well
described by the Bose-Hubbard model with Hamiltonian\cite{hubbardtoolbox,Fis89}
\begin{equation}
\hat{H} =-J\sum_{\langle i,j\rangle }{\hat{b}}_{i}^{\dag }{\hat{b}}_{j}+\frac{U%
}{2}\sum_{i}{\hat{b}}_{i}^{\dag }{\hat{b}}_{i}\left( {\hat{b}}_{i}^{\dag }{%
\hat{b}}_{i}-1\right) +\sum_{i}\epsilon _{i}{\hat{b}}_{i}^{\dag }{\hat{b}}%
_{i}.  \label{BH}
\end{equation}%

Here ${\hat{b}}_{i}$ (${\hat{b}}_{i}^{\dag }$) are destruction (creation)
operators for the bosonic atoms at site $i$. $J/\hbar$ denotes the nearest
neighbour tunnelling rate, $U$ the on-site collisional energy shift, and $
\epsilon_i$ the background potential. The high degree of control available over
the parameters in this system, e.g., changing the relative values of $U$ and
$J$ by varying the lattice depth, $V_0$, has led to seminal experiments on
\emph{strongly correlated} gases in optical lattices, e.g., the study of the
superfluid-Mott insulator transition\cite{Gre02}, the realisation of 1D quantum
liquids with atomic gases\cite{Par04,Sto04} (see also\cite{Kin04,Lab04}), and
the investigation of disordered systems\cite{Fal06}. 3D optical lattices have
also opened new avenues in cold collision physics and
chemistry\cite{Fed04,Ryu05,Sto06,Tha06}.


A striking prediction of the Bose-Hubbard Hamiltonian (\ref{BH}) is the
existence of stable repulsively bound atom pairs. These are most intuitively
understood for strong repulsive interaction $|U| \gg J$, $U>0$, where an
example of such a pair is a state of two atoms occupying a single site,
$|2_{i}\rangle \equiv ({\hat{b}}_{i}^{\dag}{}^{2}|$vac$\rangle)/\sqrt{2}$. This
state has a potential energy offset $U$ with respect to states where the atoms
are separated (see Fig. 1a). The pair is unable to decay by converting the
potential energy into kinetic energy, as the Bloch band allows a maximum
kinetic energy for two atoms given by $8\, J$, twice its width. The pair can
move around the lattice, with both atoms tunnelling to a neighbouring site
(Fig. 1b), but the atoms cannot move independently. The stability of
repulsively bound pairs is intimately connected with the absence of
dissipation, e.g., in contrast to solid state lattices, where interactions with
phonons typically lead to rapid relaxation.


 We obtain experimental evidence for repulsively bound pairs with a sample of ultracold $^{87}$Rb
atoms in a 3D optical lattice with lattice period $a=415.22\,$nm. The key tool
used to prepare and observe the pairs is their adiabatic conversion into
chemically bound dimers using a magnetic-field sweep across a Feshbach
resonance\cite{Koe06,Don02,Reg03,Her03,Xu03,Cub03,Dur04} (at 1007.40 G, see
\cite{Volz03,Tha06}). The initial state is prepared from a pure sample of
Rb$_2$ Feshbach molecules in the vibrational ground state of the lattice where
each lattice site is occupied by not more than a single molecule\cite{Tha06}.
Sweeping across the Feshbach resonance we adiabatically dissociate the dimers
and obtain a lattice correspondingly filled with $2 \times 10^4$ atom pairs, at
an effective filling factor of typically $0.3$. In all our experiments reported
here, the preparation takes place at a deep lattice of $V_0=35 \, E_r$
($E_r=2\pi^2\hbar^2/m\lambda^2$, where $m$ is the mass of the atoms),
corresponding to $J/\hbar\approx 2 \pi \times 0.7\,$Hz and $U/J = 3700$. Away
from the Feshbach resonance, the effective interaction between the atoms is
repulsive with scattering length $a_s=+100 \, a_0$ (where $a_0$ is the Bohr
radius).

\begin{figure}
\includegraphics[width=1\columnwidth]{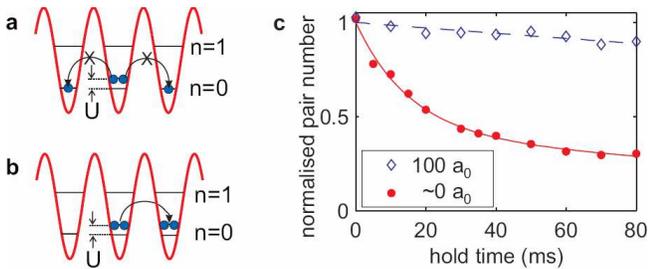}
\caption{ Atom pairs in an optical lattice. (a) Repulsive interaction
(scattering length $a>0$) between two atoms sharing a lattice site in the
lowest band (n = 0) gives rise to an interaction energy $U$. Breaking up of the
pair is suppressed due to the lattice band structure and energy conservation.
(b) The pair is a composite object that can tunnel through the lattice. (c)
Long lifetime of repulsively bound atom pairs which are held in a 3D optical
lattice. The potential depth
 is $ (10\pm 0.5)\,E_r$ in one direction and $(35\pm 1.5)\,E_r$ in the perpendicular directions.
Shown is the remaining fraction of pairs for a scattering length of
100\,a$_{0}$ (open diamonds) and a scattering length of about $(0\pm
10)$\,a$_{0}$ (filled circles) as a function of the hold time. The lines are
fit curves of an exponential (dashed line) and the sum of two exponentials
(solid line). }
\end{figure}


In order to demonstrate the stability of repulsively bound pairs, we lower the
lattice potential in one direction in 1\,ms to a depth of $V_0 = 10 \, E_r$.
This increases dramatically the tunnelling rates along this direction to
$J/\hbar\approx 2\pi \times 63\,$Hz, potentially allowing the atom pairs to
quickly separate. After a variable hold time we determine the number of
remaining pairs. This is done by adiabatically raising the lattice to its full
initial depth of $V_0 = 35\,E_r$, and converting doubly occupied sites to
Feshbach molecules with near unit efficiency\cite{Tha06}. A purification
pulse\cite{Tha06} then removes all remaining atoms due to dissociated pairs.
Afterwards the molecules are again converted back into atoms, and can then be
detected via conventional absorption imaging.

The results of these lifetime measurements are shown in Fig. 1c. For repulsive
interaction ($a_s=100\,a_{0}$) the atom pair sample exhibits the remarkably
long lifetime of $700$\,ms (exponential fit). This lifetime is mainly limited
by inelastic scattering of lattice photons\cite{Tha06}, and greatly exceeds the
calculated time for an atom to tunnel from one site to the next,
$2\pi\hbar/(4J) \sim 4\,$ms. In contrast, if we turn off the on-site
interaction by tuning the scattering length near zero, we observe a much faster
decay in the number of doubly occupied sites due to the rapid diffusion of
unbound atoms through the lattice (see Fig.~1c). This observation clearly
demonstrates that the stability of the pairs is induced by the on-site
interaction $U$.


We can more deeply understand these repulsively bound pairs through an exact solution of the two
particle Lippmann-Schwinger
scattering equation based on the Bose-Hubbard model. We write the two atom wavefunction as $%
\Psi (\mathbf{x},\mathbf{y})$, where the positions of the two particles
 are denoted $\mathbf{x}=\sum_{i}x_{i}%
\mathbf{e}_{i}$ and $\mathbf{y}=\sum_{i}y_{i}%
\mathbf{e}_{i}$, with $\mathbf{e}_{i}$ being the primitive lattice vectors, and
$x_{i},y_{i}$ integer numbers. Introducing centre of mass,
$\mathbf{R}=(\mathbf{x}+\mathbf{y})/2$, and relative coordinates, $\mathbf{r}%
=\mathbf{x}-\mathbf{y}$, we can solve the Schr\"odinger equation with the
ansatz $\Psi (\mathbf{x},\mathbf{y})=\exp (i\mathbf{K}\mathbf{R})\psi_K (%
\mathbf{r})$, where $\mathbf{K}$ is the quasi-momentum of the centre of mass
motion. We derive two types of solutions (for details see appendix
\ref{app:exactsolution}), each of which are eigenstates of the centre of mass
quasi-momentum $\mathbf{K}$. These states, as illustrated in Fig. 2a,
correspond to (i) unbound \textit{scattering solutions} (shaded area in Fig.
2a), where the two particles move on the lattice, and scatter from each other
according to the interaction $U$,
 and (ii) \textit{repulsively bound pairs} for which the pair wavefunction $\psi_{K} (\mathbf{r})$ is square
integrable. In $1D$ and $2D$, states of
 repulsively bound pairs always exist for non-zero $U$, while in $3D$ they exist
above a critical value $U_{\rm crit}\approx 0.5J$.

In this letter, we primarily focus on the 1D situation, which in the experiment
corresponds to a low depth of the lattice along one direction, whilst the
lattice in the perpendicular directions remains very deep ($35E_r$). Here the
energy of the bound pairs is $E(K)=2 J\left[\sqrt{4( \cos \frac{Ka}{2})
^{2}+(U/2J) ^2}+2\right]$. This is plotted in Fig. 2a as the Bloch band of a
stable composite object \emph{above} the continuum of two particle scattering
states. In the limit of strong interaction, $U>>J$, this reduces to $E(K) \sim
4J+U+(4J^2/U) (1+\cos Ka)$, which shows that the bound pairs indeed have
binding energy $\approx U$ and hop through the lattice with an effective
tunnelling rate $J^{2}/(\hbar U)$.

\begin{figure*}[tbp]
 \includegraphics[width=\textwidth]{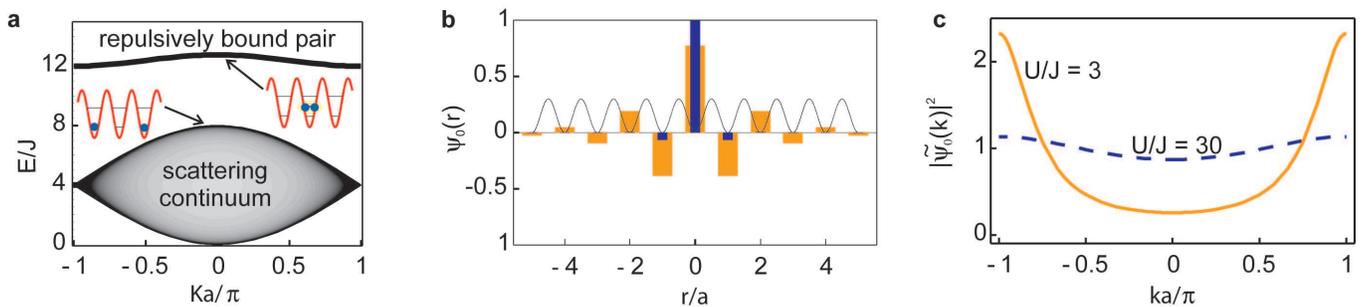}
\caption{Atom pair states in 1D. (a) Energy spectrum of the 1D Hamiltonian for
$U/J=5$ ($V_0\approx 4\,E_r$) as a function of centre of mass quasi-momentum
$K$. The Bloch band for repulsively bound pairs is located above the continuum
of unbound states. The grey level for the shading of the continuum is
proportional to the density of states. (b) The pair wavefunction $\psi_0(r)$,
showing the amplitude at each site with $U/J=30$ ($V_0\approx 10\, E_r$, narrow
bars) and $U/J=3$ ($V_0\approx 3\, E_r$, wide bars). (c) The square modulus of
the corresponding momentum space wavefunctions $|\tilde \psi_0(k)|^2$, which
are equivalent to the single particle momentum distributions, since $K=0$.
Note the characteristic peaks at the edge of the Brillouin zone.}
\end{figure*}

Figure 2b shows the pair wavefunctions $\psi_K (r)$ for repulsively bound pairs
($a_s = 100 a_0$) in 1D with $K=0$, for $U/J=30$ $(V_0 = 10 E_r)$ and $U/J=3$
$(V_0 = 3 E_r)$.
 For large $U/J$, bound pairs essentially consist
of two atoms occupying the same site, whereas for small $U/J$, the pair is
delocalised over several lattice sites. The corresponding quasi-momentum
distribution can be found from the Fourier transform $\tilde \psi_0(k)$ of the
pair wavefunction (see Fig. 2c). Since $K=0$, $|\tilde\psi_0(k)|^2$ is also
equal to the single particle quasi-momentum distribution. When the two
particles are localised on the same site, the quasi-momentum distribution is
essentially flat. However, for lower $U/J$ the wavefunction is
characteristically peaked at the edges of the Brillouin zone. This occurs
because the energy of the repulsively bound state is above that of the
continuum, and thus the contribution to the corresponding wavepacket of single
particle quasi-momentum states with higher energy is favoured.
 In contrast, if we had $U<0$, the pair would be
attractively bound, and would have energy lower than that in the continuum.
Thus contributions from the low energy quasi-momentum states would be favoured,
leading to a single peak in the centre of the Brillouin zone. In both cases,
the amplitude of the peaks grows with increasing width $4J$ of the Bloch band.
In general, the stable bound pairs will not be prepared in a fixed
quasi-momentum state $K$ in an experiment, but rather in a superposition of
different momentum states. For non-zero $K$, the peaks in the single particle
quasi-momentum distribution are translated by $K$, but their strength is also
reduced. As a consequence, for typical symmetric distributions of $K$, the peak
at the edge of the Brillouin zone remains present, however, is less strong than
the optimal case of vanishing $K$. We have verified this using many-body
numerical simulations, which are performed using time dependent density-matrix
renormalisation group methods\cite{vidal, dmrgvidaldaley, dmrgvidalwhite}.


We have experimentally investigated the quasi-momentum distribution of the
pairs in various regimes by mapping it onto a spatial distribution, which we
measured using standard absorption imaging. For this, we first adiabatically
lower the lattice depth in the $X$ - direction (see Fig. 3a) to a pre-chosen
height while the lattice depth in the other two directions are kept high (35
$E_r  $). This will prepare repulsively bound pairs at the chosen lattice
depth. We then turn off the lattice rapidly enough so that the pair
wavefunction cannot change, but slowly with respect to the bandgap, so that
single-particle quasi-momenta are mapped to real momenta\cite{Gre01,Den02}.  We
have typically employed linear ramps with rates of 0.2 $E_r / \mu$s.  The
resulting momentum distribution is converted to a spatial distribution after
$\sim$ 15 ms time of flight.

\begin{figure}
  \includegraphics[width=\columnwidth]{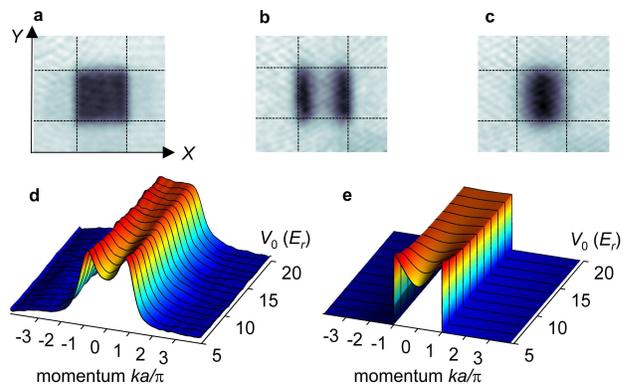}
  \caption{ Quasi-momentum distribution of atoms in the lattice.
Images (a) - (c) show absorption images of the atomic distribution after
release from the 3D lattice and a subsequent
$15\,$ms time of flight. 
The horizontal and vertical black lines
 enclose the first Brillouin zone.
(a) Distribution when lattice sites are occupied by single atoms. (b)
Distribution for repulsively bound atom pairs (see text for details). (c) Same
as (b) but pairs are attractively bound. Panels (d) [experiment] and (e)
[numerical calculation] depict the quasi-momentum distribution for pairs in the
$X$-direction as a function of lattice depth $V_0$, after integration over the
$Y$-direction.}\label{momentum}
\end{figure}

Figure 3\,(a-c) shows typical measured quasi-momentum distributions which were
obtained after adiabatically lowering the lattice depth in the $X$ - direction
to lowest lattice depths below $3\,E_r$. If only empty sites and sites with
single atoms are present in the lattice, then the first Brillouin zone is
homogenously filled\cite{Gre01} (Fig. 3\,a). For repulsively bound pairs the
momentum distribution is, in general, peaked at the edges of the first
Brillouin zone (Fig. 3\,b), whereas for attractively bound pairs it is peaked
in the center of the first Brillouin zone (Fig. 3\,c). In order to change the
interaction between the atoms from repulsive to attractive, we change the
scattering length making use of the Feshbach resonance\cite{Volz03} at 1007.40
G. Figures ~\ref{momentum}\,d and e show the dependence on lattice depth $V_0$
of the single particle quasi-momentum distribution for repulsively bound pairs
from experiment and numerical simulation, respectively. As expected, the peak
structure is more pronounced for lower values of $V_0$, and diminishes for
larger $V_0$. This characteristic is a clear signature of the pair wavefunction
for repulsively bound pairs.

We also performed spectroscopic measurements, determining the binding energy
from dissociation of the pairs by modulating the depth of the lattice at a
chosen frequency. On resonance, the modulation allows pairs to release their
binding energy. Figure 4\,a shows the number of remaining pairs as a function
of the modulation frequency. This was repeated for a variety of lattice depths
$V_0$ in one direction while keeping the lattice in the other two directions at
$35\, E_r$. The behaviour of the binding energy as a function of the lattice
depth provides an additional key signature of repulsively bound pairs. As shown
in Fig. 4b, the resonance positions are in good agreement with numerical
simulations and essentially coincide with interaction energy, $U$.

\begin{figure}
  \includegraphics[width=\columnwidth]{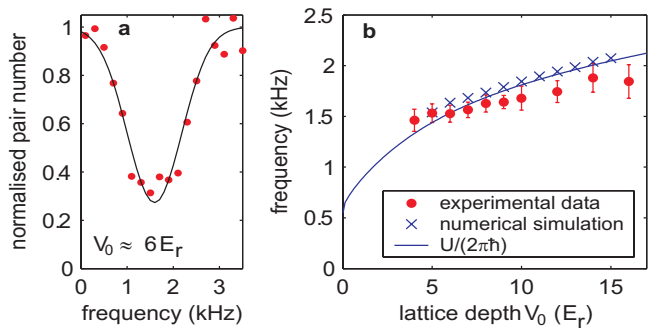}
  \caption{Modulation spectroscopy of repulsively bound pairs.
  (a)  Typical resonance dip showing the remaining
 number of atom pairs as a function of the modulation frequency,
  for $V_0\approx 6\,E_r$.
   The black line is a Gaussian fit, a choice which was justified by numerical calculations.
   (b) Plot showing the measured resonance frequencies (filled circles) as a
   function of the lattice depth. They show good agreement with numerical simulations
   (crosses) and also coincide with the onsite collisional energy shift $U$ (line).
   Experimental error bars correspond to the 95\% confidence interval for the Gaussian
   fit parameters of the resonance
   dips.}
\label{modulation}
\end{figure}

It is important to note that for sufficiently large $U/J$, \textit{repulsively
bound pairs} are stable under collisions with each other. This is particularly
evident in the limit $U>>J$ where, by energy arguments, the elastic scattering
between pairs is the only open channel. This means that even a relatively dense
quantum lattice gas of these objects can be long-lived. When the lattice height
is lowered so that $U/J$ becomes sufficiently small, it is possible for a
certain fraction of the pairs to dissociate by collision with other pairs. In
our experiments, we observe the onset of this behaviour for lattice depths
lower than 6\,$E_r$, i.e., $U/J \sim 9$. The dynamics of the collisions and
details of the decay depend crucially on lattice depth and the local density of
pairs across the lattice. Further details of these processes will be discussed
elsewhere.

In conclusion, we have demonstrated the formation of a novel composite object
in an optical lattice: a stable bound state that arises from the lattice band
structure and repulsion between the constituents. Although no direct analog to
repulsively bound atomic pairs is known to exist, the formation of a metastable
state is reminiscent of trapping light in photonic bandgap materials
\cite{photonicCrystal}, or extended lifetimes of excited atoms in cavity
quantum electrodynamics\cite{Cavity}. In both cases, decay is suppressed by
restriction of the accessible light field modes. Stable repulsively bound
objects should be viewed as a general phenomenon and their existence will be
ubiquitous in cold atoms lattice physics. They give rise also to new potential
composites with fermions\cite{Hof02} or Bose-Fermi mixtures\cite{Lew04}, and
can be formed in an analogous manner with more than two particles. The
stability of repulsively bound objects could thus be the basis of a wealth of
new quantum many body states or phases.

\appendix
\section{Exact solution for single pair bound state}
\label{app:exactsolution}
 Within the Bose-Hubbard Model (1) the Schr\"odinger
equation describing two particles in a homogenous optical lattice takes the
form
\begin{equation}
\left[ -J \left(\tilde{\Delta}^{0}_{{\bf x}} + \tilde{\Delta}_{{\bf y}}^{0}
\right) +  U \delta_{{\bf x},{\bf y}}\right] \Psi({\bf x},{\bf y}) = E \:
\Psi({\bf x},{\bf y}), \label{schroedinger}
\end{equation}
where the vectors ${\bf x}$ and  ${\bf y}$ describe the positions of the two
particles as defined in the main text. The operator $\tilde{\Delta}^{{\bf
K}}_{{\bf x}} \Psi({\bf x})\!=\! \sum_{i=1}^d \cos\left({\bf K}{\bf
e}_{i}/2\right)\left[ \Psi({\bf x\!+\!e}_{i}) \!+\! \Psi({\bf x\!-\!e}_{i}) - 2
\Psi({\bf x})\right]$ denotes the discrete lattice Laplacian with $d$ the
dimensionality in the cubic lattice, and $\delta_{{\bf x},{\bf y}}$ is a
Kronecker delta. Writing the wavefunction in relative and centre of mass
coordinates $\Psi({\bf x}, {\bf y}) = \exp(i {\bf K} {\bf R}) \psi_{K}({\bf
r})$, the Schr\"odinger equation (\ref{schroedinger}) then reduces to a single
particle problem in the relative coordinate
\begin{equation}
  \left[ -2 J \tilde{\Delta}^{\bf K}_{\bf r}  + E_{\bs K}
  + U \delta_{{\bf r},0}\right]
\psi_K({\bf r}) = E \psi_K({\bf r}) \label{singleparticleSE}
\end{equation}
with $E_{\bs K}= 4 J\sum_{i} \left[1\!-\!\cos ({\bf K}{\bf e}_{i}/2) \right]$
being the kinetic energy of the centre of mass motion.

The short range character of the interaction potential allows for a resummation
of the perturbation expansion generated by the corresponding Lippmann-Schwinger
equation. We obtain the scattering states
\begin{equation}
 \psi^{(+)}({\bf r}) = \exp( i {\bf k} {\bf r}) - 8 \pi J f_E({\bf K}) G_{{\bs K}}(E, {\bf r}) \label{scatteringstates}
\end{equation}
with scattering amplitude
\begin{equation}
f_E({\bf K})= - \frac{1}{4\pi}\frac{ U/(2J) }{1-G_{\bs K}(E,0) U}
\end{equation}
where the total energy $E=  \epsilon_{{\bs k},{\bs K}} +  E_{{\bs K}}$, and
$\epsilon_{{\bs k},{\bs K}}= 4 J \sum_{i} \cos({\bf K}{\bf e}_{i}
/2)\left[1-\cos({\bf k}{\bf e}_{i})\right]$. Furthermore, $G_{{\bs K}}(E,{\bf
r})$ denotes the Greens function of the non-interacting problem, which in
Fourier space takes the form $\tilde G_{\bf K}(E,{\bf k}) =1/(E- \epsilon_{{\bs
k},{\bs K}} + i \eta)$.

The energy spectrum for these states in one dimension is shown as a function of
$K$ by the shaded region in Fig.~2a. In addition, the pole in the scattering
amplitude indicates the presence of an additional bound state. The energy
$E_{\rs bs}$ of the bound state is determined by $G_{\bs K}(E_{\rs bs},0) U=1$
and the bound state wavefunction takes the form $\psi^{\rs bs}({\bf r}) = c \:
G_{\bs K}(E_{\rs bs},{\bf r})$ with $c$ being a normalisation factor.

\begin{acknowledgments} The authors thank Helmut Ritsch for helpful
discussions, and Matthias Theis and Stefan Schmid for their help in setting up
the experiment. We acknowledge support from the Austrian Science Fund (FWF)
within the Spezialforschungsbereich 15, from the European Union within the
OLAQUI and SCALA networks, from the TMR network "Cold Molecules" under contract
No. HPRN-CT-2002-00290, and the Tiroler Zukunftsstiftung.
\end{acknowledgments}



\begin{thebibliography}{99}

\bibitem{hubbardtoolbox}  Jaksch, D. \& Zoller, P.
The cold atom Hubbard toolbox. Annals of Physics \textbf{315}, 52-79
(2005), and references therein.

\bibitem{Blo05}
Bloch, I.
Ultracold quantum gases in optical lattices. Nature
Physics \textbf{1}, 23-30 (2005).


 \bibitem{Fis89}
 Fisher, M. P. A., Weichman, P. B., Grinstein, G. \& Fisher, D. S.
 Boson localization and the superfluid insulator transition. Phys.
 Rev. B \textbf{40}, 546-570 (1989).

\bibitem{Gre02}
Greiner, M., Mandel, O., Esslinger, T., H\"ansch, T. W. \& Bloch, I.
Quantum phase transition from a superfluid to a Mott insulator in a
gas of ultracold atoms. Nature \textbf{415}, 39-44 (2002).

\bibitem{Par04}
Paredes, B. {\it et al.}
Tonks - Girardeau gas of ultracold atoms in an optical lattice.
 Nature \textbf{429},  277-281 (2004).

\bibitem{Sto04}
 Stöferle, T.,   Moritz, H.,  Schori, C.,  Köhl, M. \&   Esslinger, T.
Transition from a strongly interacting 1D superfluid to a Mott
insulator. Phys. Rev. Lett. \textbf{92}, 130403 (2004).

\bibitem{Kin04}
Kinoshita, T., Wenger, T. \& S. Weiss, D. S.
 Observation of a one-dimensional Tonks-Girardeau gas.  Science
 \textbf{305}, 1125-1128 (2004).

\bibitem{Lab04}
 Laburthe Tolra, B. \textit{et al.}
Observation of reduced three-body recombination in a correlated 1D
degenerate Bose gas. Phys. Rev. Lett. \textbf{92}, 190401 (2004).



\bibitem{Fal06}
 Fallani, L.,  Lye, J. E., Guarrera, V.,  Fort, C. \&  Inguscio, M.
   Onset of a Bose-Glass of ultracold atoms in a
disordered crystal of light. Preprint at
(http://arxiv.org/abs/cond-mat/0603655) (2006).



\bibitem{Fed04}
Fedichev, P. O., Bijlsma, M. J. \& Zoller, P. Extended molecules and geometric
scattering resonances in optical lattices. Phys. Rev. Lett. \textbf{92}, 080401
(2004).


\bibitem{Ryu05}
Ryu, C., \textit{et al.}
Raman-induced oscillation between an atomic and a molecular quantum gas.
Preprint at (http://arxiv.org/abs/cond-mat/0508201) (2005).

\bibitem{Sto06}
St\"oferle, T., Moritz, H., G\"unter, K., K\"ohl, M., \& Esslinger,
T., Molecules of fermionic atoms in an optical lattice. Phys. Rev.
Lett. \textbf{96}, 030401 (2006).

\bibitem{Tha06}  Thalhammer, G., 
 Winkler, K.,   Lang, F.,  Schmid, S.,  Grimm, R.
 \&  Hecker Denschlag, J.
 Long-lived Feshbach molecules in a 3D optical lattice.
Phys. Rev. Lett.  \textbf{96}, 050402 (2006).



\bibitem{Don02}
Donley, E. A., Claussen, N. R., Thompson, S. T. \& Wieman, C. E.
Atom - molecule
 coherence in a Bose-Einstein condensate. Nature \textbf{417}, 529-533 (2002).

\bibitem{Reg03} 
 Regal, C. A., Ticknor, C.,   Bohn, J. L. \&   Jin, D. S.
Creation of ultracold molecules from a Fermi gas of atoms. Nature
\textbf{424}, 47-50 (2003).

\bibitem{Her03}
J. Herbig \textit{et al.} Preparation of a pure molecular quantum
gas. Science \textbf{301}, 1510 - 1513 (2003).

\bibitem{Xu03}
 Xu, K. \textit{et al.}
 Formation of quantum-degenerate sodium molecules.
 Phys. Rev. Lett. \textbf{91}, 210402 (2003).




\bibitem{Cub03}
 Cubizolles, J., Bourdel, T.,  Kokkelmans, S. J. J. M. F.,  Shlyapnikov, G. V. \&
   Salomon, C.
Production of long-lived ultracold Li$_2$ molecules from a Fermi
gas. Phys. Rev. Lett. \textbf{91}, 240401 (2003).

\bibitem{Dur04}
D\"urr, S., Volz, T., Marte A. \& Rempe, G. Observation of molecules
produced from a Bose-Einstein condensate.
 Phys. Rev. Lett. \textbf{92}, 020406 (2004).

\bibitem{Koe06}
Koehler, T., Goral, K. \& Julienne, P.S. Production of cold
molecules via magnetically tunable Feshbach resonances. Preprint at
(http://arxiv.org/abs/cond-mat/0601420) (2006).

\bibitem{Volz03} Volz, T., Dürr, S., Ernst, S., Marte, A. \&  Rempe, G.
Characterization of elastic scattering near a Feshbach resonance in $^{87}$Rb.
Physical Review A \textbf{68}, 010702 (2003).



\bibitem{vidal} Vidal, G.
Efficient classical simulation of slightly entangled quantum
computations. Phys. Rev. Lett. \textbf{91}, 147902 (2003).


\bibitem{dmrgvidaldaley}
Daley, A. J.,  Kollath, C.,
 Schollw\"{o}ck, U. \& Vidal, G.
Time-dependent density-matrix renormalization-group using adaptive effective
Hilbert spaces.
 J. Stat. Mech.: Theor. Exp. P04005, 1-28
(2004).

\bibitem{dmrgvidalwhite}
 White, S.R. \&  Feiguin, A.E.
  Real-time evolution using the density
matrix renormalization Group.
 Phys. Rev. Lett. \textbf{93}, 076401 (2004).





\bibitem{Gre01}
 Greiner, M.,  Bloch, I.,  Mandel, O.,  H\"ansch, T.~W., \& Esslinger,
 T. Exploring phase coherence in a 2D lattice of Bose-Einstein
condensates. Phys. Rev. Lett. \textbf{87}, 160405 (2001).

\bibitem{Den02}  Hecker Denschlag, J. \textit{et al.}
A Bose-Einstein condensate in an optical lattice.
 J. Phys. B  \textbf{35}, 3095-3110 (2002).



\bibitem{photonicCrystal}
 Joannopoulos, J. D.,  Meade, R. D. \&  Winn, J. N.
 \textit{Photonic Crystals: Molding the Flow of Light}
(Princeton Univ. Press, Princeton, September 1995).

\bibitem{Cavity}
 Berman, P. (Ed.)
\textit{Cavity Quantum Electrodynamics } (Academic Press, New York,
1994).



\bibitem{Hof02}
Hofstetter, W., Cirac, J. I., Zoller, P., Demler, E. \& Lukin, M. D.
High-temperature superfluidity of fermionic atoms in optical
lattices. Phys. Rev. Lett. \textbf{89}, 220407 (2002).

\bibitem{Lew04}
Lewenstein, M., Santos, L., Baranov, M. A. \& Fehrmann, H. Atomic
Bose-Fermi mixtures in an optical lattice. Phys. Rev. Lett.
\textbf{92}, 050401 (2004).












\end{thebibliography}
\end{document}